%% file: main_draft.tex
\documentclass[prd,twocolumn,showpacs,amsmath,amssymb,nofootinbib]{revtex4-2}
\usepackage[utf8]{inputenc}
\usepackage{booktabs}
\usepackage{feynmp}
\usepackage{epsf}
\usepackage{graphicx}
\usepackage{epsfig}
\usepackage{xcolor}
\usepackage{subfigure}
\usepackage{pstricks}
\usepackage{pst-node}
\usepackage{rotating}
\usepackage{graphics}
\usepackage{latexsym}
\usepackage{dcolumn}
\usepackage{bm}
\usepackage{times}
\usepackage{indentfirst}
\usepackage{float}
\usepackage{overpic}
\usepackage{amsmath}
\usepackage{appendix}
\usepackage{upgreek}
\usepackage{chngcntr}
\usepackage{amssymb}%
\usepackage{pifont}%
\usepackage[top=0.5in,bottom=0.8in,left=0.9in,right=0.9in]{geometry}

\usepackage{subfigure}
\usepackage{lineno}

\usepackage{multirow}
\usepackage{makecell}

\usepackage{hyperref}
\hypersetup{
	colorlinks=true,
	linkcolor=blue,
	filecolor=blue,
	urlcolor=blue,
	citecolor=blue,
}

\newcommand{\jpsi}{J/\psi}
\newcommand{\dz}{D^{0}}
\newcommand{\dzbar}{\bar{D}^{0}}

\newcommand{\pp}{\pi^+\pi^-}
\newcommand{\ppp}{\pi^+\pi^-\pi^0}
\newcommand{\pip}{\pi^+}
\newcommand{\pim}{\pi^-}
\newcommand{\piz}{\pi^0}

\newcommand{\Mppp}{\it M_{\pi^{+}\pi^{-}\pi^{0}}}

\newcommand{\Mpp}{\it M_{\it{\pi^+\pi^-}}}
\newcommand{\Mkshort}{\it M_{\it{K_{S}^{0}}}}

\newcommand{\kaonp}{K^+}

\newcommand{\kshort}{K_{S}^{0}}
\newcommand{\klong}{K_{L}^{0}}
\newcommand{\ksl}{K_{S,L}^0}
\newcommand{\ee}{e^+e^-}
\newcommand{\DeltaE}{\Delta {\it E}}
\newcommand{\Mbc}{M_{\rm BC}}
\newcommand{\Ebeam}{E_{\rm beam}}
\newcommand{\DzbartoKp}{\bar{D}^0\to K^+\pi^-}
\newcommand{\DzbartoKpp}{\bar{D}^0\to K^+\pi^-\pi^0}
\newcommand{\DzbartoKppp}{\bar{D}^0\to K^+\pi^-\pi^+\pi^-}
\newcommand{\DDbar}{D\bar{D}}
\newcommand{\DzDzbar}{D^0\bar{D}^0}
\newcommand{\pmiss}{\vec{p}_{\rm miss}}
\newcommand{\emiss}{E_{\rm miss}}
\newcommand{\mmiss}{M_{\rm miss}}
\newcommand{\NST}{N^{\rm ST}}
\newcommand{\NSig}{N^{\rm sig}}
\newcommand{\NSG}{N^{\rm SG}}
\newcommand{\NSB}{N^{\rm SB}}
\newcommand{\NNET}{N^{\rm NET}}

\def\TeV{\ifmmode {\mathrm{\ Te\kern -0.1em V}}\else
                   \textrm{Te\kern -0.1em V}\fi}%
\def\GeV{\ifmmode {\mathrm{\ Ge\kern -0.1em V}}\else
                   \textrm{Ge\kern -0.1em V}\fi}%
\def\MeV{\ifmmode {\mathrm{\ Me\kern -0.1em V}}\else
                   \textrm{Me\kern -0.1em V}\fi}%
\def\keV{\ifmmode {\mathrm{\ ke\kern -0.1em V}}\else
                   \textrm{ke\kern -0.1em V}\fi}%
\def\eV{\ifmmode  {\mathrm{\ e\kern -0.1em V}}\else
                   \textrm{e\kern -0.1em V}\fi}%

\let\gev=\GeV

\def\TeVc{\ifmmode {\mathrm{\ Te\kern -0.1em V}/c}\else
                   {\textrm{Te\kern -0.1em V}/$c$}\fi}%
\def\GeVc{\ifmmode {\mathrm{\ Ge\kern -0.1em V}/c}\else
                   {\textrm{Ge\kern -0.1em V}/$c$}\fi}%
\def\MeVc{\ifmmode {\mathrm{\ Me\kern -0.1em V}/c}\else
                   {\textrm{Me\kern -0.1em V}/$c$}\fi}%
\def\keVc{\ifmmode {\mathrm{\ ke\kern -0.1em V}/c}\else
                   {\textrm{ke\kern -0.1em V}/$c$}\fi}%
\def\eVc{\ifmmode  {\mathrm{\ e\kern -0.1em V}/c}\else
                   {\textrm{e\kern -0.1em V}/$c$}\fi}%

\let\gevc=\GeVc

\def\TeVcc{\ifmmode {\mathrm{\ Te\kern -0.1em V}/c^2}\else
                   {\textrm{Te\kern -0.1em V}/$c^2$}\fi}%
\def\GeVcc{\ifmmode {\mathrm{\ Ge\kern -0.1em V}/c^2}\else
                   {\textrm{Ge\kern -0.1em V}/$c^2$}\fi}%
\def\MeVcc{\ifmmode {\mathrm{\ Me\kern -0.1em V}/c^2}\else
                   {\textrm{Me\kern -0.1em V}/$c^2$}\fi}%
\def\keVcc{\ifmmode {\mathrm{\ ke\kern -0.1em V}/c^2}\else
                   {\textrm{ke\kern -0.1em V}/$c^2$}\fi}%
\def\eVcc{\ifmmode  {\mathrm{\ e\kern -0.1em V}/c^2}\else
                   {\textrm{e\kern -0.1em V}/$c^2$}\fi}%

\let\gevcc=\GeVcc

\def\cm{\ifmmode  {\mathrm{\ cm}}\else
                   \textrm{~cm}\fi}%

\def\ifb{\mbox{fb$^{-1}$}}

		\newcommand{\bfg}{\begin{figure}}
			\newcommand{\efg}{\end{figure}}
		\newcommand{\bitm}{\begin{itemize}}
			\newcommand{\eitm}{\end{itemize}}
		\newcommand{\bnum}{\begin{enumerate}}
			\newcommand{\enum}{\end{enumerate}}
		\newcommand{\btbl}{\begin{table}}
			\newcommand{\etbl}{\end{table}}
		\newcommand{\btbu}{\begin{tabular}}
			\newcommand{\etbu}{\end{tabular}}
		\newcommand{\bcl}{\begin{center}}
			\newcommand{\ecl}{\end{center}}
		
		\newcommand{\beq}{\begin{equation}}
			\newcommand{\eeq}{\end{equation}}
		\newcommand{\beqr}{\begin{eqnarray}}
			\newcommand{\eeqr}{\end{eqnarray}}

\newcommand{\BESIIIorcid}[1]{\href{https://orcid.org/#1}{\hspace*{0.1em}\raisebox{-0.45ex}{\includegraphics[width=1em]{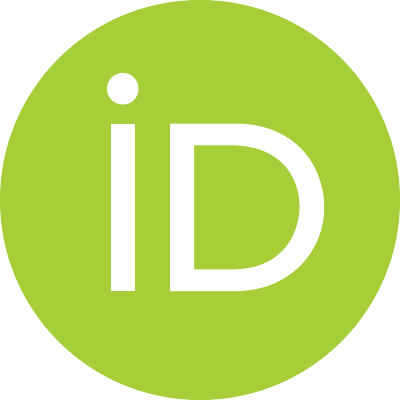}}}} 
\begin{document}
	
\title{\boldmath Study of $\kshort$-$\klong$ asymmetry in the decays $\dz\to\kshort\omega $ and $\dz\to\klong\omega $}

\author{
\input{authorlist_2025-11-25}
}
\date{September 3, 2026}
\renewcommand{\arraystretch}{1.8}
\renewcommand{\abstractname}{}

\begin{abstract}
	Based on $\ee$ annihilation data corresponding to an integrated luminosity of 7.93~$\ifb$ collected at a center-of-mass energy of 3.773~$\gev$ with the BESIII detector at the BEPCII collider, the absolute branching fractions of the decays $\dz \to \kshort \omega$ and $\dz \to \klong\omega$ are measured to be $(11.79 \pm 0.19 \pm 0.26 \pm 0.47) \times 10^{-3}$ and ($10.84 \pm 0.14 \pm 0.23 \pm 0.44) \times 10^{-3}$, respectively. 
    The $\kshort$-$\klong$ branching-fraction asymmetry of these two decays is  $R(\dz,\ksl \omega) = \frac{\mathcal{B}(\dz\to\kshort\omega) - \mathcal{B}(\dz \to\klong\omega )}{\mathcal{B}(\dz \to\kshort\omega) + \mathcal{B}(\dz\to\klong\omega)} =(4.2 \pm 1.0 \pm 0.9 \pm 2.8)\%$.
    Here, the first uncertainties are statistical, the second systematic, and the third arise from the interference between $\dz \to \ksl\omega$ and the non-resonant $\dz \to \ppp\ksl$ processes.
\end{abstract}
\maketitle
\section{INTRODUCTION}
\label{sec:intro}
 
Two-body non-leptonic decays of charm mesons provide a useful platform to study their underlying decay mechanism, test the Kobayashi-Maskawa (KM) mechanism in the Standard Model (SM), and search for new physics beyond the SM~\cite{ARU.58.249}. 
However, a theoretical description of the underlying decay mechanism based on a QCD-inspired approach for exclusive non-leptonic charm meson decays is challenging. 
This is related to the charm-quark mass of order 1.5~$\gevcc$, which is neither heavy enough to allow for a sensible heavy-quark mass expansion nor light enough for an application of chiral perturbation theory. 
Nevertheless, a model-independent approach to charm meson non-leptonic decays based on the topological diagram approach (TDA)~\cite{PRD.81.074021} is available. 
This approach classifies the decay amplitudes in terms of six distinct diagrams according to the topology of weak interactions, with all strong interaction effects included, and enables the extraction of the individual topological amplitudes while disentangling different decay mechanisms.
Similarly, the factorization-assisted topological-amplitude (FAT) approach~\cite{PRD.86.036012, PRD.89.054006} combines the topological-amplitude parametrization with the conventional naive factorization hypothesis for splitting the short-distance and long-distance dynamics into Wilson coefficients and hadronic matrix elements of four-fermion operators (topological amplitudes). 
Both of the above approaches require  sufficient and precise experimental measurements as inputs.
The decays $D\to PV$, where $P$ and $V$ denote pseudo-scalar and vector mesons, respectively, are of particular interest due to their sizable branching fractions (BFs) in charm meson decays, and have been studied intensively within the frameworks of the TDA~\cite{PRD.93.114010, PRD.100.093002, PRD.109.073008} and FAT~\cite{PRD.86.036012, PRD.89.054006} approaches.
In these studies, the BFs of the Cabibbo-favored (CF) processes, such as $D\to \bar{K^0}V$, are the important inputs to determine the topological amplitudes and the corresponding phase. 

The $\kshort-\klong$ asymmetry in the two-body charm meson decays $D \to\ksl X$,  defined as
\begin{equation}
\label{eq:Rdef}
	R(D,\ksl X) = \frac{\mathcal{B}(D \to \kshort X) - \mathcal{B}(D \to \klong X)}{\mathcal{B}(D \to \kshort X) + \mathcal{B}(D \to \klong X)},
\end{equation}
was first proposed in Ref.~\cite{PLB.349.363} to describe the difference in BFs of charm meson decays between modes with $\kshort$ and $\klong$ involved in the final states. 
Non-zero values of $R(D,\ksl X)$ are induced by the interference between the CF decay ($D\to \bar{K}^0 X$) and  the doubly Cabibbo-suppressed (DCS) decay ($D\to K^0 X$), giving access to DCS processes.
Precise measurements of $R(D,\ksl X)$ enable us to determine the amplitude of the DCS processes, which are essential to examine whether the topological amplitudes in DCS decays are the same as those in CF decays in the TDA framework~\cite{PRD.109.073008}. Furthermore, these measurements test flavor SU(3) symmetry, probe the dynamics of charm hadron decays, and provide understanding of the $\dz-\dzbar$ mixing mechanism~\cite{PRD.81.114020, PRD.80.076008, PRD.86.014014, PRD.92.014004, PRD.95.073007, CPC.42.063101, PRD.109.073008}.

Two-body non-leptonic decays $D\to PP$ and $D\to PV$ have been the subject of many measurements, especially the CF and singly Cabibbo-suppressed modes~\cite{pdg}. 
In addition to precise BFs for CF processes, $\kshort-\klong$ asymmetries for decays with final-state $\ksl$ are also measured, and are very useful for theoretical work. 
In general, the theoretical predictions from the TDA~\cite{PRD.109.073008} and FAT~\cite{PRD.95.073007} models agree with the data well. However, some discrepancies remain, particularly for $R(D,\ksl X)$ in the decays $\dz\to \ksl V$ with $V=\omega/\phi$.
Both theoretical models predict the BF relation $\mathcal{B}(\dz \to \kshort V)>\mathcal{B}(\dz \to \klong V)$, which is contrary to experimental measurements.
The latest measurement from the BESIII experiment reported $R(\dz,\ksl \omega) = -0.024\pm0.031$~\cite{KLX}, which deviates from the FAT prediction by 4.4$\sigma$\cite{PRD.95.073007} and from the TDA prediction by more than 2$\sigma$~\cite{PRD.109.073008}.
This measurement is obtained by measuring the BF of $\dz\to \klong \omega$ with the early BESIII  data collected at a center-of-mass energy of 3.773~GeV corresponding to an integrated luminosity 2.93~$\ifb$, and then calculating $R(\dz,\ksl \omega)$ with the BF of  $\dz\to \kshort \omega$ reported by the Particle Data Group (PDG)~\cite{pdg2022}.
Therefore, new experimental measurements are required to clarify this tension and improve our understanding of the relevant decay mechanisms.

In this paper, we update the measurements of the absolute BFs of $\dz \to \kshort\omega$ and $\dz \to \klong\omega$ and extracted the corresponding $\kshort-\klong$ asymmetry  $R(\dz,\ksl \omega)$, using the double-tag (DT) method~\cite{MARK-III:1989deaDTMethod}, which is implemented to improve the ratio of signal to background and to minimize the systematic uncertainties. In the DT method, a $\dzbar$ meson is first reconstructed with its tag decay modes; this is referred to as a single tag (ST).  Next, the $\dz$ meson is reconstructed with signal modes $\dz\to\ksl\omega$ in the selected ST subsample; those cases where candidates are found are referred to as DTs.
Measurements are performed using a data sample with an integrated luminosity of 7.93~$\ifb$, collected at a center-of-mass energy of 3.773~GeV by the BESIII detector operating at the BEPCII collider.

Throughout this paper, charge conjugation is always implied unless explicitly noted otherwise.

\section{BESIII DETECTOR AND MONTE-CARLO SIMULATION}
\label{sec:detmc}
The BESIII detector~\cite{Ablikim:2009aa} records the symmetric $\ee$ collision events provided by the BEPCII storage ring~\cite{Yu:IPAC2016-TUYA01} with center-of-mass energies ($\sqrt{s}$) ranging from 1.84 to 4.95~$\gev$. BEPCII achieves a peak luminosity of $1.1 \times10^{33}\;\text{cm}^{-2}\text{s}^{-1}$  at $\sqrt{s} = 3.773$~$\gev$.
BESIII has collected large data samples in this energy region~\cite{Ablikim:2019hff}. 
The cylindrical core of BESIII detector covers 93\% of the full solid angle and consists of a helium-based multilayer drift chamber~(MDC), a plastic scintillator time-of-flight system~(TOF), and a CsI(Tl) electromagnetic calorimeter~(EMC), which are all enclosed in a superconducting solenoidal magnet providing a 1.0-T magnetic field. 
The solenoid is supported by an octagonal flux-return yoke with resistive plate counter muon identification modules interleaved with steel.
The momentum resolution of charged particle at 1~$\gevc$ is $0.5\%$, and the specific ionization energy loss~(d$E$/d$x$) resolution is $6\%$ for electrons from Bhabha scattering events.
The EMC measures photon energies with a resolution of $2.5\%$ ($5\%$) at 1~$\gev$ in the barrel (endcap) region. 
The time resolution in the TOF barrel region is 68~ps, while that in the endcap region was 110~ps. 
The endcap TOF system was upgraded in 2015 using multi-gap resistive plate chamber
technology~\cite{tof1,tof2,tof3}, providing a time resolution of 60~ps; 63\% of the data used in this analysis has this improved resolution.

Monte Carlo (MC) simulated data samples produced with a {\sc geant4}-based~\cite{geant4} software package, which includes the geometric description of the BESIII detector and the detector response, are used to determine detection efficiencies and to estimate backgrounds. 
The effects of beam-energy spread and initial-state radiation (ISR) in the $\ee$ annihilation are modelled with the generator {\sc kkmc}~\cite{ref:kkmc}. 
The inclusive MC sample includes the production of $D\bar{D}$ pairs (including quantum coherence for the neutral $D$ channels), non-$D\bar{D}$ decays of the $\psi(3770)$, ISR production of the $\jpsi$ and $\psi(3686)$ states, and continuum processes. 
All particle decays are modelled with {\sc evtgen}~\cite{ref:evtgenl,ref:evtgenp} using BFs 
either from the PDG~\cite{pdg} when available,
or otherwise from {\sc lundcharm}~\cite{ref:lundcharm1,ref:lundcharm2}.

Final-state radiation from charged particles is incorporated using the {\sc photos} package~\cite{photos}.
The signal samples of $\dz\to \kshort\omega$ and $\dz\to \klong\omega$ with the subsequent decays $\omega \to \ppp$ and $\kshort \to \pp$ are generated with {\sc evtgen}~\cite{ref:evtgenl,ref:evtgenp}, where the decays $\dz\to \kshort \omega$ and $\dz\to \klong\omega$ are described with the SVS model~\cite{ref:evtgenp}, the decay $\omega\to\ppp$ with a model based on a Dalitz-plot analysis~\cite{PhysRevD.98.112007}, 
and the decays $\kshort \to \pp$ and  $\piz \to \gamma\gamma$ with phase space.

\section{EVENT SELECTION}
\label{EvtSel}

\subsection{Selection of final-state particles}
\label{sub: comsel}

Tracks detected in the MDC are required to be within a polar angle ($\theta$) range of $|\rm{cos\theta}|<0.93$, where $\theta$ is defined with respect to the $z$-axis, which is the symmetry axis of the MDC.  For tracks not originating from $K_S^0$ decay, the distance of closest approach to the interaction point (IP)  must be less than 10\,cm along the $z$-axis, $|V_{z}|$, and less than 1\,cm in the transverse plane, $|V_{xy}|$.
Charged particle identification~(PID) of tracks is performed by combining the d$E$/d$x$ measured by the MDC and the flight time measured by the TOF to form the likelihood values $\mathcal{L}(h)$ for $h=K$ and $\pi$ hypotheses.
Charged kaons and pions are identified by requiring $\mathcal{L}(K)>\mathcal{L}(\pi)$ and $\mathcal{L}(\pi)>\mathcal{L}(K)$, respectively.

Photon candidates are identified using isolated showers in the EMC. The deposited energy of each shower must be more than 25~MeV in the barrel region ($|\cos \theta|< 0.80$) and more than 50~MeV in the end-cap region ($0.86 <|\cos \theta|< 0.92$). To exclude showers that originate from charged tracks, the angle subtended by the EMC shower and the position of the closest charged track at the EMC must be greater than 10 degrees, as measured from the IP. 

To suppress fake showers from electronic noise and showers unrelated to the event, the difference between the EMC time and the event start time is required to be within 
[0, 700]\,ns.

The $\piz$ candidates are reconstructed from pairs of photon candidates that have an invariant mass within the interval (0.115, 0.150)~\gevcc.
To improve momentum resolution, a kinematic fit is performed by constraining the $\gamma\gamma$ invariant mass to the $\piz$ nominal mass~\cite{pdg} with fit quality satisfying $\chi^{2}<50$; the resultant kinematic variables are used for subsequent analysis.

The $K_{S}^0$ candidates are reconstructed from two tracks associated with opposite charge that satisfy $|V_{z}|<$ 20~cm. The tracks are assigned the pion-mass hypothesis without imposing PID criteria. They are constrained to originate from a common vertex and are required to have an invariant mass
within $|M_{\pi^{+}\pi^{-}} - m_{K_{S}^{0}}|<$ 12~MeV$/c^{2}$, where $m_{K_{S}^{0}}$ is the $K^0_{S}$ nominal mass~\cite{pdg}. The decay length of the $K^0_S$ candidate is required to be greater than twice the vertex resolution away from the IP. 

\subsection{Selection of single tag candidates}
\label{sub:stsel}

Candidate $\ee\to \dz\dzbar$ events are selected from those preselected  by the ST method, which reconstructs a $\dzbar$ meson using three hadronic decay modes with large decay BFs and relatively low background: $\DzbartoKp$, $\DzbartoKpp$, and $\DzbartoKppp$. 

The identification of ST candidates uses two kinematic variables the energy difference with respect to the beam energy, $\DeltaE$, and the beam-energy-constrained mass, $\Mbc$. These variables are defined as
\begin{equation}
\label{eq1}
\begin{aligned}
    \DeltaE^{\rm {tag}} & = {\it{E}_{\it \dzbar}} - \Ebeam,  \\
    \Mbc^{\rm {tag}}    & = \sqrt{\Ebeam^{2}/\it{c}^{\rm 4} - \rm\vec{p}_{\it\dzbar}^{\rm 2}/\it{c}^{\rm 2}},
\end{aligned}
\end{equation}
\noindent
where $\Ebeam$ is the beam energy, and ${\it E_{\it \dzbar}}$ ($\it\vec{p}_{\it \dzbar}$) is the reconstructed energy (momentum) of ST $\dzbar$ candidate in the center-of-mass frame of $\ee$ system. 
In the case of multiple $\dzbar$ candidates for a specific ST mode in an event, only the one with the smallest $|\DeltaE^{\rm tag}|$ is chosen for further analysis. 

To reduce the peaking background from $\dzbar \to \kshort \kaonp \pim$ in the ST mode $\DzbartoKppp$, events with ${|\Mpp - \Mkshort| <}$~30 MeV$/c^{2}$ are rejected, where ${\Mkshort}$ is the known $\kshort$ mass \cite{pdg}. 
For the ST mode $\DzbartoKp$, backgrounds from cosmic rays and Bhabha events are rejected using the same criteria described in Ref.~\cite{BESIII:2014rtmDTagT}.
To further reject combinational backgrounds, $\dzbar$ candidates are required to have $\DeltaE^{\rm tag}$ within an interval corresponding to three times the resolution, as specified in Table~\ref{tab:deST}.
\begin{table}[t!]
	\centering
	\caption{Summary of $\Delta E^{\rm tag}$ requirements for the three ST modes.}
	\begin{tabular}{ l|c }
		\hline
		\hline
		~~~~Decay Modes~~~~ & ~~~~$\Delta \rm{E}$ $(\gev)$~~~~ \\
		\hline
		~~~~$\bar{D}^{0} \to K^{+} \pi^{-}$ & (-0.027, 0.027)  \\
		~~~~$\bar{D}^{0} \to K^{+} \pi^{-} \pi^{0}$ & (-0.062, 0.049) \\
		~~~~$\bar{D}^{0} \to K^{+} \pi^{-} \pi^{+} \pi^{-}$~~~~ &~~~~ (-0.026, 0.024)~~~~ \\
		\hline
		\hline
	\end{tabular}
	\label{tab:deST}
\end{table}
\subsection{Selection of double-tag candidates}
\label{sub:dtsel}
After ST selection, the DT candidate events are selected by reconstructing a $\dz$ signal recoiling against ST $\dzbar$ candidates. 

The signal candidates for $\dz \to \kshort\omega$ decays are reconstructed from a $\kshort$ and $\ppp$ combination made from particles that are unused in the ST reconstruction. 
The variables $\DeltaE^{\rm sig}$ and $\Mbc^{\rm sig}$, defined similarly to Eq.~(\ref{eq1}), are used to identify signal $\dz$ candidates. 
If there are multiple $\dz$ candidates, the one with the smallest $|\DeltaE^{\rm sig}|$ is retained for subsequent analysis. 
A requirement of $-0.055 <\DeltaE^{\rm sig} < 0.040\gev$ is applied to reduce backgrounds from the $\ee\to q\bar{q}$ continuum and from $\ee\to D\bar{D}$ mis-reconstruction. 

The signal candidates for $\dz \to \klong\omega$ are reconstructed using the missing mass method. 
In addition to the reconstructed particles used in the ST side, the candidate events are further required to have another $\pp$ pair without any extra charged tracks and at least one $\piz$ candidate. 
If multiple $\piz$ candidates exist, the one with the minimum $\chi^{2}$ of the kinematic fit for $\piz \to \gamma\gamma$ is retained for further analysis.
The $\klong$ recoils against the combination of the ST $\dzbar$ with the selected $\ppp$ candidate, and is identified using the missing-mass squared variable, $\mmiss^{2}$, which is defined as

\begin{equation}
	\begin{aligned}
		\mmiss^{2}   & = \emiss^{2}/{\it c}^4 - \pmiss^{2}/{\it c}^2, \\
		\emiss       & = \Ebeam - E_{\ppp}, \\
		\pmiss       & = \vec{\it p}_{\dz} - \vec{\it p}_{\ppp},
	\end{aligned}
\label{eq3}
\end{equation}
\noindent
where $\emiss$ and $\pmiss$ are the energy and momentum of the missing particle, 
${\it E}_{\ppp}$ and $\vec{\it p}_{\ppp}$ are the corresponding variables of the $\ppp$ system.
To improve the momentum and mass resolution, in Eq.~(\ref{eq3}) the $\dz$ energy is assigned to be the beam energy $\Ebeam$, and its momentum $\vec{\it p}_{\dz}$ is opposite to that of ST $\dzbar$ in direction and with the magnitude expected for the decay $\psi(3770) \to \dz\dzbar$. 

To avoid misidentification of $\klong$-induced showers in the EMC as photons originating from  $\piz$ decay, photon candidates must be separated by more than $15^{\circ}$ from the $\klong$ trajectory.  
 To suppress dominant peaking backgrounds from $\dz \to \kshort \omega$, with subsequent decay $\kshort \to \piz\piz$, and $\dz \to \eta \omega$, with subsequent decay $\eta \to \gamma\gamma$, the total energy of extra photons situated away from the $\klong$ direction with an opening angle greater than $15^{\circ}$, $E_{\gamma}^{\rm extra}$ must be less than 0.27~$\gev$. 
 
\section{BACKGROUND STUDIES AND SIGNAL YIELD EXTRACTION}
\label{SignalYields}

Based on the above selection criteria, the potential backgrounds for the ST and DT samples are studied with data and MC samples.
The signal yields of ST ($\NST$) and DT ($\NSig$) are extracted by performing unbinned maximum likelihood fits to the corresponding distributions for selected candidates.

\subsection{Single-tag sample}

Backgrounds in the ST sample are dominated by combinatorics from $\ee\to q\bar{q}$ and $\ee \to \DDbar$ events. 
Detailed studies of the ST signal MC samples indicate that a small fraction of selected candidates are incorrect combinations of $\dzbar$ decay final states, resulting in a broad shape in the  $\Mbc^{\rm tag}$ distribution. 
\begin{figure}[ht!]
	\begin{center}
		\subfigure{\includegraphics[width=0.45\textwidth]{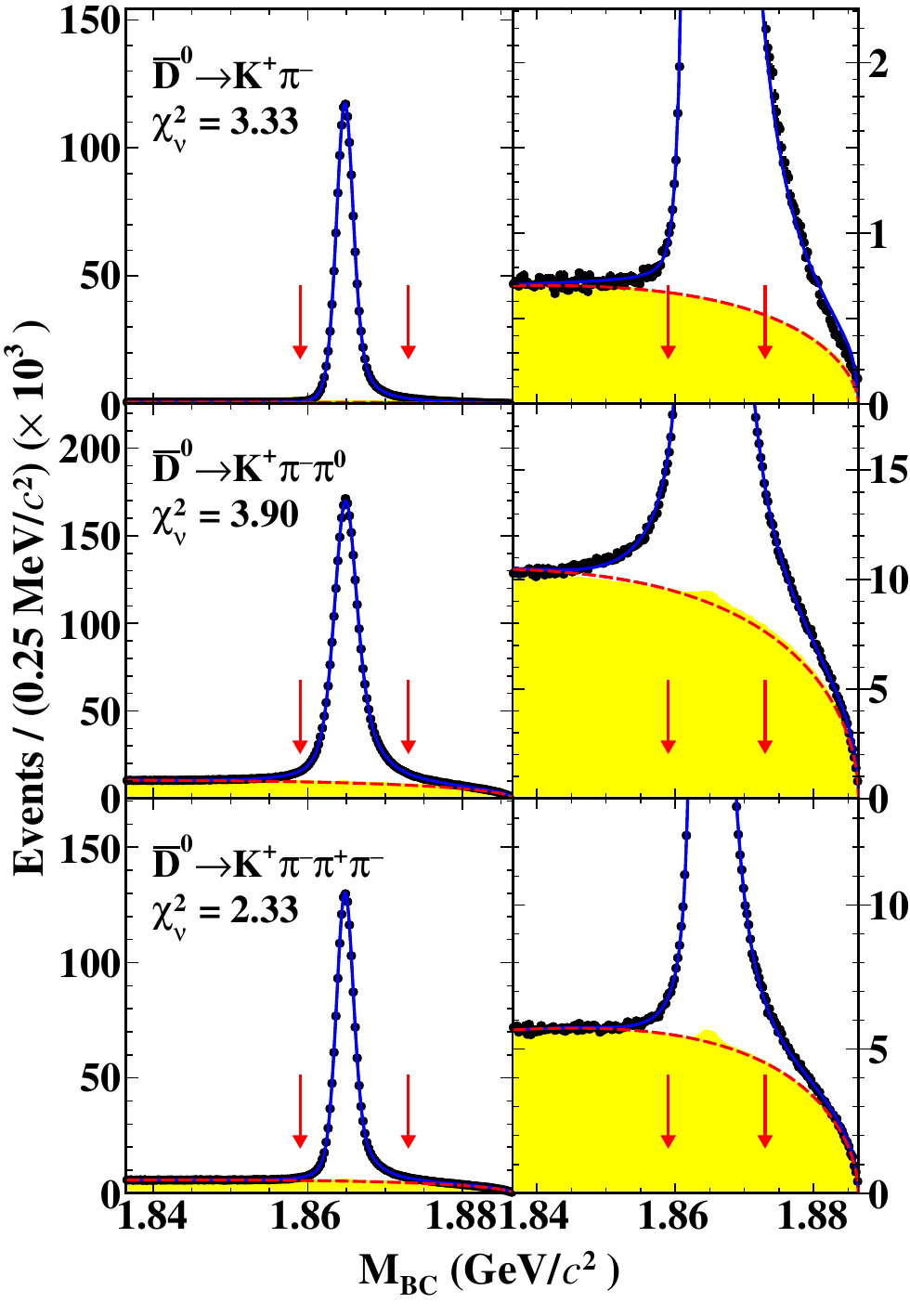}}
        \abovecaptionskip=0pt
		\caption{Fits to the $\Mbc^{\rm tag}$ distributions of the candidates for the three ST modes. The points with error bars represent data. The solid blue curves depict the fit results, the dashed red curves and yellow histograms are the backgrounds from the fits and the inclusive MC sample, respectively. The interval between the two red arrows indicates the $\Mbc^{\rm tag}$ signal region for the $\dz\to \klong\omega$ reconstruction. The right plots are with the large zoom to depict the background shapes. The goodness of fit is quantified by the reduced chi-square, $\chi^2_\nu=\chi^2/{\rm n.d.f}.$, where the ${\rm n.d.f.} = 191$ is number of degrees of freedom of the fit for the three ST modes. }
		\label{fig:tagmbcfit}
	\end{center}
\end{figure}

The $\Mbc^{\rm tag}$ distributions are shown in Fig.~\ref{fig:tagmbcfit} for the three ST modes individually. 
Maximum-likelihood fits on the $\Mbc^{\rm tag}$ distributions are used to extract the ST yields.
In the fits, the signal is modelled with an MC-simulated shape convolved with a double Gaussian function, which accounts for the resolution difference between data and MC simulation.
The background is described by the ARGUS function~\cite{Arguscite} with a fixed cut-off value at 1.8865~$\gevcc$.
In the fits, incorrect combinatorial ST signal candidates are regarded as backgrounds. Therefore, the signal shape is extracted with ST signal MC-simulated events with an opening angle between the true and reconstructed momenta less than $15^{\circ}$ for each daughter particle.

The fit curves are shown in Fig.~\ref{fig:tagmbcfit}, which demonstrates the excellent agreement between the data distributions and the fit curves.
The fitted ST yields and the corresponding detection efficiencies are summarized in Table~\ref{tab:NfitST}.
The ST detection efficiencies are estimated with the same event selection and fitting procedure applied to the inclusive  MC sample, and the ratios of the fitted to produced signal yields are taken as the efficiencies.
In the analysis of $\dz\to \klong\omega$, an additional requirement ${\rm 1.859<\Mbc^{\rm tag}<1.873}~\gevcc$ is imposed to suppress the combinatorial background. The corresponding ST yields in data and efficiencies, including the impact of this requirement, are extracted and summarized in Table~\ref{tab:NfitST}.

\begin{table}[t!]
	\centering
	\caption{Summary of ST yields in data and detection efficiencies in the full $\Mbc^{\rm tag}$ region and in the restricted range $1.859<\Mbc^{\rm tag}<1.873~\gevcc$ for the three ST modes; uncertainties are statistical only. The former and latter $N^{\rm ST}, \epsilon^{\rm ST}$ numbers are used for the $\dz\to\kshort\omega$ and $\dz\to\klong\omega$ analyses, respectively.}
	\begin{tabular}{ c|c c c }
		\hline
		\hline
		ST mode        & $K^-\pip$ & $K^-\pip\piz$ &  $K^{-} \pi^{+} \pi^{-} \pi^{+}$ \\
		\hline  
     & \multicolumn{3}{c}{Full $\Mbc^{\rm tag}$ region} \\
        $N^{\rm ST}(10^3)$        & $1491.0\pm1.3$ & $3123.7\pm2.1$ & $1693.2\pm1.5$ \\
		$\epsilon^{\rm ST}$(\%) & $67.00\pm0.01$ & $37.66\pm0.01$ & $36.29\pm0.01$ \\  \hline
        & \multicolumn{3}{c}{$1.859<\Mbc^{\rm tag}<1.873~\gevcc$} \\ 
		$N^{\rm ST}(10^3)$        & $1457.3\pm1.3$ & $2918.4\pm2.0$ & $1645.6\pm1.5$ \\
        $\epsilon^{\rm ST}$(\%) & $65.39\pm0.01$ & $35.58\pm0.01$ & $35.37\pm0.01$\\
		\hline \hline
	\end{tabular}
	\label{tab:NfitST}
\end{table}

\subsection{DT $\dz \to \kshort\omega$ sample}
\label{subdztoksomega}

Candidate $\dz\to \kshort \omega$ decays are reconstructed from a $\ppp$ combination and a $\kshort$ candidate.
The corresponding distributions of the $\ppp$ invariant mass $\Mppp$ are shown in Fig.~\ref{fig:fitM3piplots} for the data and the inclusive MC sample, where an $\omega$ signal is observed over a smooth background; good agreement is seen between the data and the inclusive MC sample distributions.
The process $\dz \to \kshort \ppp$ excluding the $\omega$-intermediate state is an irreducible background.
In the following analysis, the $\omega$ signal region is defined as $0.752<\Mppp$$< 0.812$~\gevcc, which corresponds to three times the mass resolution, while the $\Mppp$-sideband region is defined as $(0.662, 0.722) \cup (0.842, 0.902)~\gevcc$.

\begin{figure}[htbp]
    \centering
    \includegraphics[width=1.0\linewidth]{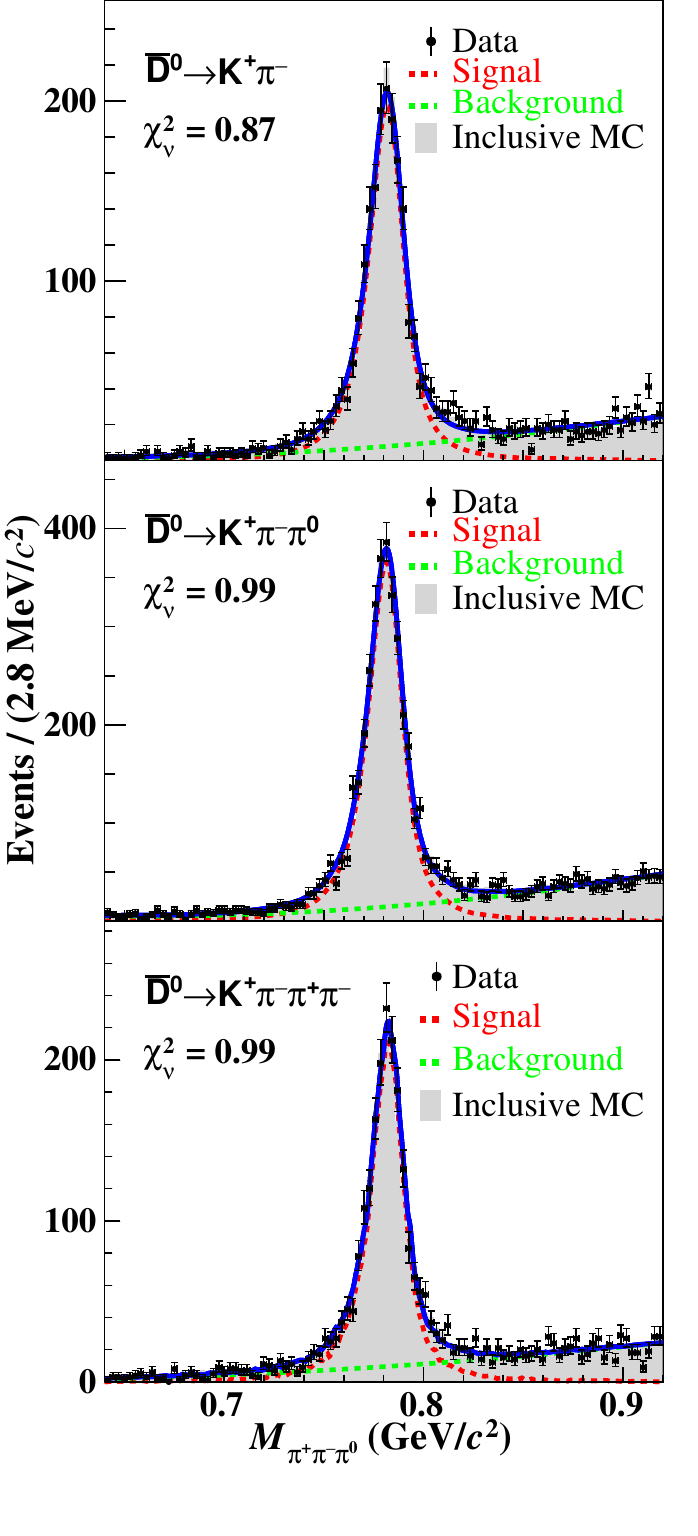}        
    \abovecaptionskip=-20pt
    \caption{Fits to the $\Mppp$ distributions for $\dz\to\kshort\omega$ DT candidates, separated by ST mode. The dots with error bars are data and the grey histograms are the inclusive MC simulation. The solid blue, dashed red and dashed magenta curves are the total fit results, signals and backgrounds, respectively. The goodness of fit is quantified by the reduced chi-square, $\chi^2_\nu=\chi^2/{\rm n.d.f.}$, where the ${\rm n.d.f.} = 94$ is number of degrees of freedom of the fits for the three ST modes.
    }
    \label{fig:fitM3piplots}
\end{figure}

The two-dimensional (2D) distributions of $\Mbc^{\rm tag}$ versus $\Mbc^{\rm sig}$ for selected events in the $\omega$ signal region are shown in Fig.~\ref{fig:2dmbsplots}. Horizontal and vertical bands around the known $\dz$ mass represent backgrounds from $\ee \to \DzDzbar$ with the incorrectly reconstructed signal $\dz$ and ST $\dzbar$, respectively. The incorrectly reconstructed signal background is referred to as BKGI and the incorrectly reconstructed ST is referred to as BKGII.
In addition, there is a diagonal band mainly from $\ee\to q\bar{q}$ background and a small fraction of signal events with both signal and ST candidates incorrectly reconstructed; this background is referred to as BKGIII.  Detailed studies based on the MC simulation show that BKGI, BKGII, and BKGIII do not produce peaks in the distribution of the incorrectly reconstructed variable.
Note that the backgrounds of  $\ee \to \DzDzbar$ with incorrectly reconstructed $\piz$ in the subsequent decay $\dzbar \to \kaonp \pim \piz$ or $\dz\to\kshort \ppp$, referred to as BKGIV and BKGV, produce a relatively broad peak in the corresponding $\Mbc^{\rm tag}$ and $\Mbc^{\rm sig}$ distributions, respectively.
\begin{figure*}[!htbp]
    \centering
    \includegraphics[width=1.0\linewidth]{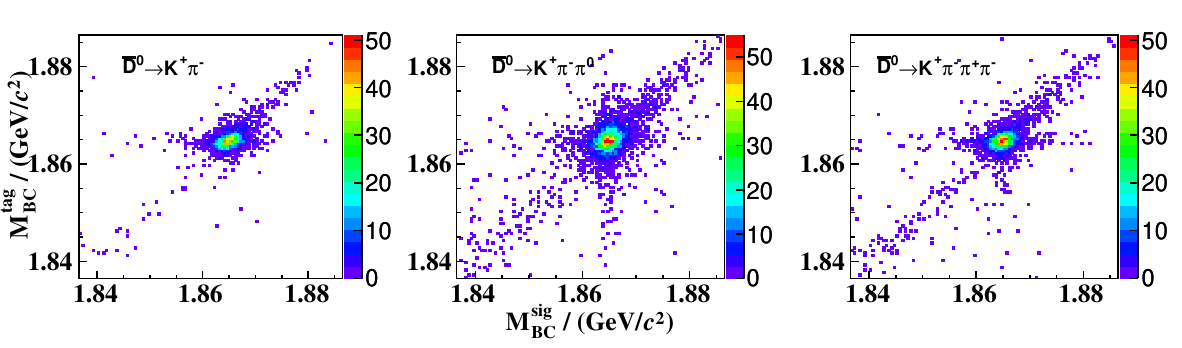}        
    \abovecaptionskip=0pt
    \caption{The $\Mbc^{\rm tag}$ versus $\Mbc^{\rm sig}$ distributions for the three ST modes.}
    \label{fig:2dmbsplots}
\end{figure*}

The DT signal yield ($\NSig$) for the decay $\dz \to \kshort \omega$ is extracted by performing a 2D unbinned maximum-likelihood fit to the distribution of $\Mbc^{\rm sig}(\equiv x_1)$ versus $\Mbc^{\rm tag}(\equiv x_2)$ of events within the $\omega$ signal region.
To estimate the irreducible background of $\dz\to \kshort\ppp$, an identical fit is performed for events in the $\omega$ sideband region.

The background yields in the $\omega$ signal region are estimated from the $\omega$ sideband region, using a scale factor, $f_{\rm scal}$, that accounts for the expected ratio of the $\omega$ signal and sideband region yields. 
The $f_{\rm scal}$ values, summarized in Table~\ref{tab:ksres} for the three ST modes individually, are obtained by fitting the $\Mppp$ distributions, as shown in Fig.~\ref{fig:fitM3piplots}. In the fit, the $\omega$ signal shape is described by the MC-simulated shape convolved with a Gaussian resolution function, and the background is described by a quadratic function.

In the above 2D maximum-likelihood fits, the signal probability density function (PDF) is
\begin{equation*}
f_{\rm SIG}=S(x_1,x_2) \otimes G(x_1, \mu_{x_1}, \sigma_{x_1}) \otimes G(x_2, \mu_{x_2}, \sigma_{x_2}),
\end{equation*}
\noindent
where $S(x_1,x_2)$ is the signal MC-simulated shape for candidates with an opening angle between the true and reconstructed momenta less than 15$^{\circ}$ for all the daughter particles, and the Gaussian functions $G(x, \mu, \sigma)$ represent the resolution difference between data and the MC simulation. In the fit, the $\mu$ and $\sigma$ parameters are fixed to those from the one-dimensional (1D) fits to the $\Mbc^{\rm sig}$ and $\Mbc^{\rm tag}$ distributions, respectively.

The PDFs of BKGI and BKGII are 
\begin{equation*}
\begin{aligned}
     f_{\rm BKGI,II} &= A(x_{i}; m_{x_{i}}, z_{x_{i}}, \rho_{x_{i}}) \\
                 &\times S(x_{j}) \otimes G(x_{j}, \mu_{x_{j}}, \sigma_{x_{j}}),
\end{aligned}
\end{equation*}
\noindent
where $i=1,2$ for BKGI and BKGII, $j=2,1$ for BKGI and BKGII. Here $A$ is an ARGUS function, \(A(x; m, z, \rho) =  x \left( 1 - \frac{x^2}{m^2} \right)^\rho e^{z\left(1 - \frac{x^2}{m^2}\right)}\) that represents the incorrectly reconstructed variable $x_1$ or $x_2$, $S(x_{2,1})$ is the projection of $S(x_1,x_2)$ for the correctly reconstructed variable, and $G(x, \mu, \sigma)$ is the same Gaussian function used to parameterize $f_{\rm SIG}$. 
In this fit, the parameter $m_{x_{1,2}}$ is fixed to 1.8865~$\gevcc$, $\rho_{x_{1,2}}$ is fixed to the values obtained from the fits to the inclusive MC sample, and $z_{x_{1,2}}$ is a free parameter.

The BKGIII PDF is 
\begin{equation*}
\begin{aligned}
     f_{\rm BKGIII} & =\it{T(x_{1}-x_{2};\mu,\sigma(x_1+x_2),n)}  \\
               & \times A(\it{x_{1}};m_{x_1},z_{x_1}^{\prime},\rho_{x_1}^{\prime})  
                 \times A(\it{x_{2}};m_{x_2},z_{x_2}^{\prime},\rho_{x_2}^{\prime}),
\end{aligned}
\end{equation*}
\noindent
where $T$ denotes the Student's $t$-distribution 
\begin{equation*}
\begin{aligned}
T(x; \mu, \sigma, n) = \frac{\Gamma(n/2 + 0.5)}{\sigma \sqrt{n\pi} \Gamma(n/2)} \left[1 + \frac{1}{2}\left(\frac{x - \mu}{\sigma}\right)^2\right]^{-\frac{n+1}{2}}\;.
\end{aligned}
\end{equation*}
\noindent
Here $\sigma$ is a function of \(x_{1} + x_{2}\), specifically \(\sigma(x_1 + x_2) = \sigma_{0} + \sigma_{1}(x_1 + x_2 - m_{x_1} - m_{x_2})\). The cut-off parameter \(m\) of $A$ is fixed at 1.8865~$\gevcc$, while other parameters are floated.

The BKGIV and BKGV PDFs are
\begin{equation*}
	f_{\rm BKGIV,V}=B_{1,2}(x_{1}, x_{2}) \otimes G(x_{1,2}; \mu_{x_{1,2}}, \sigma_{x_{1,2}}),
\end{equation*}
\noindent
where $B_{1,2}(x_{1}, x_{2})$ are the signal MC-simulated shapes with incorrectly reconstructed $\piz$, and $G(x,\mu,\sigma)$ is the same Gaussian function used to parameterize $f_{\rm SIG}$.

\begin{figure}[htp]
	\begin{center}
		\subfigure{\includegraphics[width=0.5\textwidth]{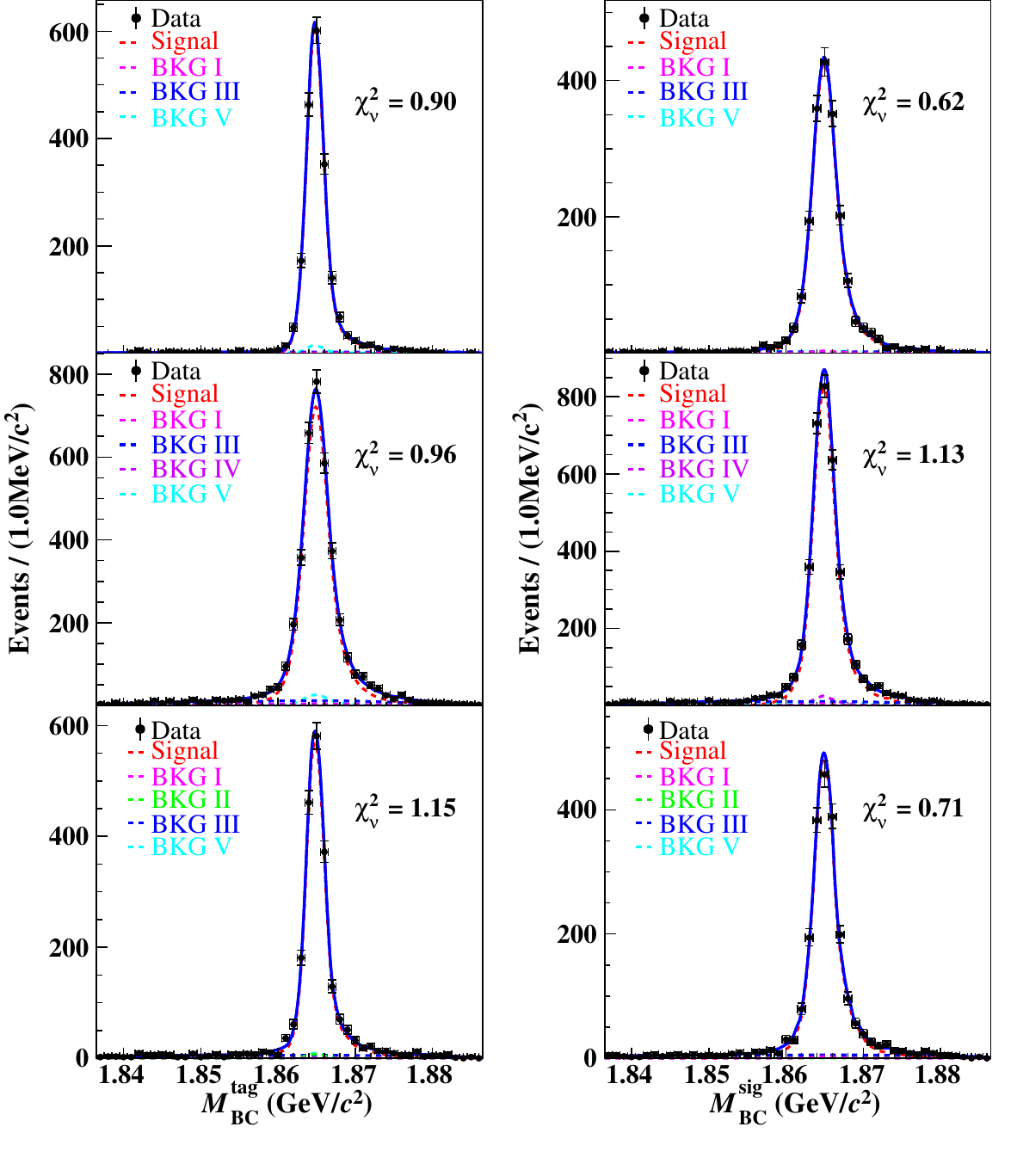}}        \abovecaptionskip=-10pt
		\caption{Distributions of $\Mbc^{\rm tag}$ and $\Mbc^{\rm sig}$ for the three ST modes: (top) $\DzbartoKp$, (middle) $\DzbartoKpp$, and (bottom) $\DzbartoKppp$ in the $\omega$ signal region. The dots with error bars are data, the solid blue curves represent the total fit results, the dashed red curves are the signal shapes, and the dashed magenta, green, blue, violet, and cyan curves represent the BKGI, BKGII, BKGIII, BKGIV, and BKGV, respectively. The goodness of fit is quantified by the reduced chi-square, $\chi^2_\nu=\chi^2/{\rm n.d.f.}$, where the ${\rm n.d.f.} = 37, 36, 37$ are numbers of degrees of freedom of the fits for the three ST modes $\DzbartoKp$, $\DzbartoKpp$, and $\DzbartoKppp$. The zoomed plots showing the specifics of background components can be found in the supplemental material.}
		\label{fig:sigmbc2d}
	\end{center}
\end{figure}
\begin{figure}[htp]
	\begin{center}
		\subfigure{\includegraphics[width=0.5\textwidth]{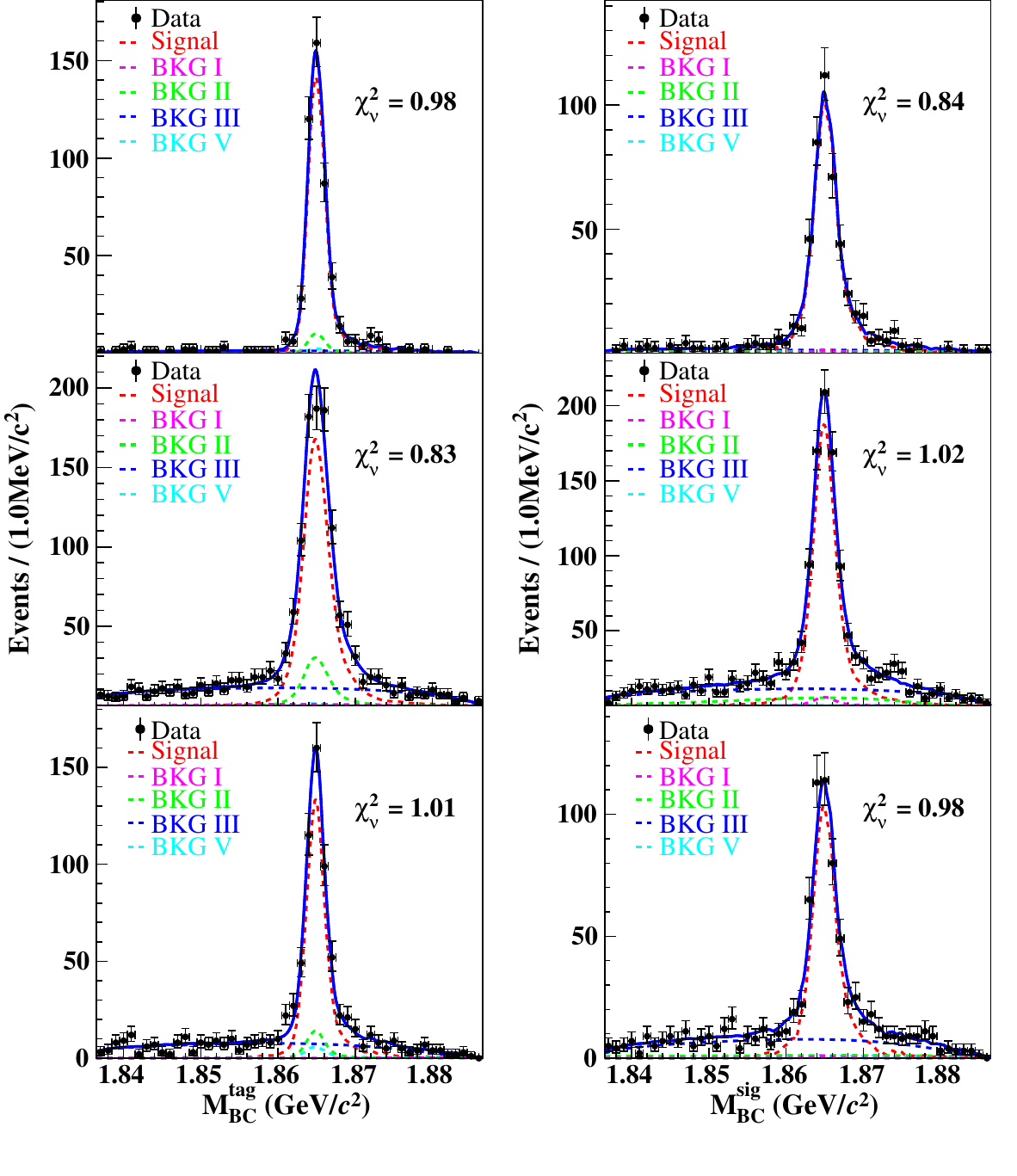}}        \abovecaptionskip=-10pt
		\caption{Distributions of $\Mbc^{\rm tag}$ and $\Mbc^{\rm sig}$ for the
  three ST modes: (top) $\DzbartoKp$, (middle) $\DzbartoKpp$, and (bottom) $\DzbartoKppp$ in the $\omega$ sideband region. The dots with error bars are data, the solid blue curves represent the total fit results, the dashed red curves are the signal shapes, and the dashed magenta, green, blue, violet, and cyan curves represent the BKGI, BKGII, BKGIII, BKGIV, and BKGV, respectively. The goodness of fit is quantified by the reduced chi-square, $\chi^2_\nu=\chi^2/{\rm n.d.f.}$, where the ${\rm n.d.f.} = 35$ is number of degrees of freedom of the fits for the three ST modes. The zoomed plots showing the specifics of background components can be found in the supplemental material.}
		\label{fig:sigmbc2dd}
	\end{center}
\end{figure}

To determine the signal and background yields, two maximum likelihood fits are performed separately on samples of selected events in the $\omega$ signal and sideband regions. The projections of these fits on the $\Mbc^{\rm tag}$ and $\Mbc^{\rm sig}$ distributions are shown in Figs.~\ref{fig:sigmbc2d} and~\ref{fig:sigmbc2dd}. 
The yields of $\dz\to \kshort \ppp$ in the $\omega$ signal region ($\NSG$) and sideband region ($\NSB$) are summarized in Table~\ref{tab:ksres}.
The net signal yield of $\dz\to \kshort\omega$ ($\NNET$) is obtained by subtracting the non-$\omega$ contribution, which is estimated from the yield in the $\omega$ sideband region, from the yield of $\dz\to \kshort\ppp$ in the $\omega$ signal region
\begin{equation}
{\NNET}={\NSG}- f_{\rm scal} \, \NSB.
\end{equation}
The detection efficiencies, $\epsilon^{\rm DT}$, are extracted using the same event selection and fitting procedure applied to signal MC samples for the three ST modes individually, which are also summarized in Table~\ref{tab:ksres}.

\begin{table}[htpb]
	\centering
        \caption{Summary of yields $\NSG$, $\NSB$, $\NNET$, scale factor $f_{\rm scal}$, and efficiencies $\epsilon^{\rm DT}$ for the three ST modes of the signal decay $D^0 \to \omega K_{S}^{0}$; uncertainties are statistical only.}
	\begin{tabular}{ c|c c c }
		\hline
		\hline
		ST Modes & ~~~$K^{+} \pim$~~~ & ~~~$K^{+} \pim\piz$~~~ & ~~~$K^{+} \pim \pip \pim$~~~ \\ \hline
        $\NSG$           & $1894.8\pm47.7$&$3406.5\pm67.1$&$1950.4\pm47.$5\\
        $\NSB$           & $435.4\pm22.4$ & $792.2\pm34.7$ & $455.6\pm24.5$\\
        $f_{\rm scal}$           & $0.356\pm0.041$ & $0.347\pm0.032$ & $0.392\pm0.046$ \\ \hline
        $\NNET$           & $1739.7\pm51.6$&$3131.4\pm72.3$&$1771.9\pm52.7$\\
        $\epsilon^{\rm DT}(\%)$ & $12.06\pm0.04$ & $5.78\pm0.03$& $5.54\pm0.02$\\
		\hline
		
		\hline
		\hline
	\end{tabular}
	\label{tab:ksres}
\end{table}

\subsection{DT $\dz \to \klong \omega$ Sample}
\label{subdztoklomega}

Candidate $\dz\to\klong\omega$ decays are reconstructed with a $\ppp$ combination and a $\klong$ candidate, where the $\klong$ is inferred by the kinematic variable $\mmiss^2$. . 
Therefore, the DT yields of $\dz\to\klong\omega$ are extracted by performing an unbinned maximum likelihood fit to the $\mmiss^2$ distributions of candidate events. Fits are performed separately in the $\omega$ signal region and sideband region; the latter fit estimates the irreducible background from $\dz\to \klong\ppp$ without the $\omega$-intermediate state. 

The $\Mppp$ distributions are shown in Fig.~\ref{fig:fitklM3piplots} for data and inclusive MC samples; good agreement is observed between the samples.
The ratios $f_{\rm scal}$ of the $\dz\to\klong\ppp$ yields in the $\omega$ signal region to those in the $\omega$ sideband region, as summarized in Table~\ref{tab:klres}, are obtained with the same approach as described in Sec.~\ref{subdztoksomega}.
\begin{figure}[htbp]
    \centering
    \includegraphics[width=1.0\linewidth]{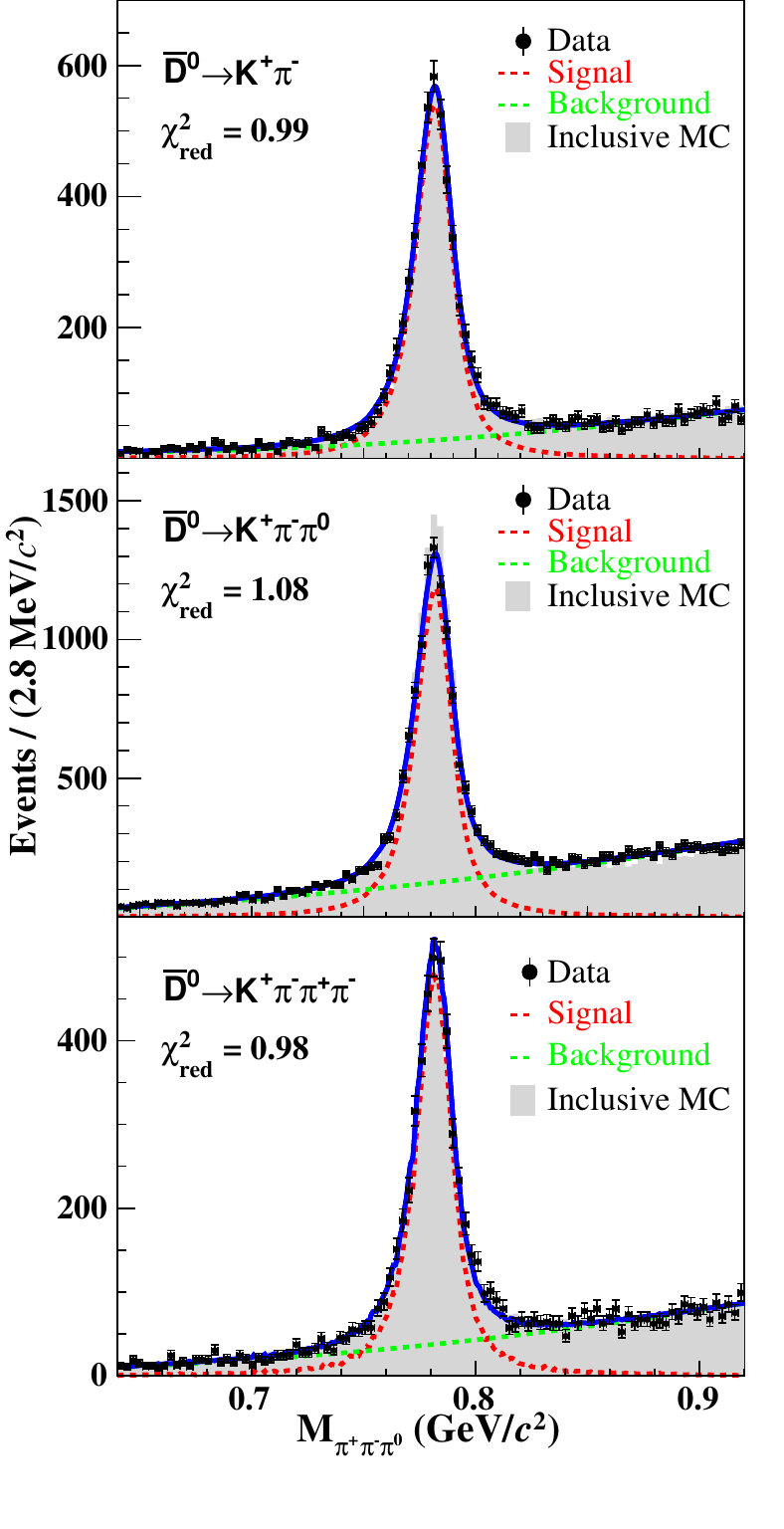}
    \abovecaptionskip=-20pt
    \caption{Fits to the $\Mppp$ distributions of $\dz\to\klong\omega$ DT candidates from the three ST modes. The dots with error bars are data, and the grey histograms are the inclusive MC simulated shape. 
    The blue solid curves are the fit results, the red dashed and magenta dashed are the signal and background, respectively. The goodness of fit is quantified by the reduced chi-square, $\chi^2_\nu=\chi^2/{\rm n.d.f.}$, where the ${\rm n.d.f.} = 94$ is number of degrees of freedom of the fits for the three ST modes.}
    \label{fig:fitklM3piplots}
\end{figure}

The $\mmiss^2$ distributions of the selected events in the $\omega$ signal and sideband regions are shown in Fig.~\ref{fig:mm2fitsig}. We observe significant $\klong$ signals and consistency between the data distribution and inclusive MC simulation. 
Assuming all events originate from $\dz$ meson decays, there are peaking backgrounds from $\dz\to \eta \omega$ and $\dz\to\kshort \omega$ in the $\mmiss^2$ distributions.
In fits to the $\mmiss^2$ distributions, the PDFs of the $\dz\to\klong \omega$ signal and  the $\dz\to\eta \omega$ background are the corresponding MC simulated shapes.
The contribution of the $\dz\to\kshort\omega$ background is not considered in the fit; instead it will be subtracted from the resulting signal yields. 
To take into account any resolution difference between data and MC simulation, the signal and background PDFs are convolved with a Gaussian function.
A cubic PDF with all parameters free describes the combinatorial background. 
Figure~\ref{fig:mm2fitsig} shows the fit results in signal and sideband regions.  The corresponding signal yields, $\NSG$ and $\NSB$, are summarized in Table~\ref{tab:klres}.

\begin{figure*}[htbp]
	\begin{center}
		\subfigure{\includegraphics[width=0.4\textwidth]{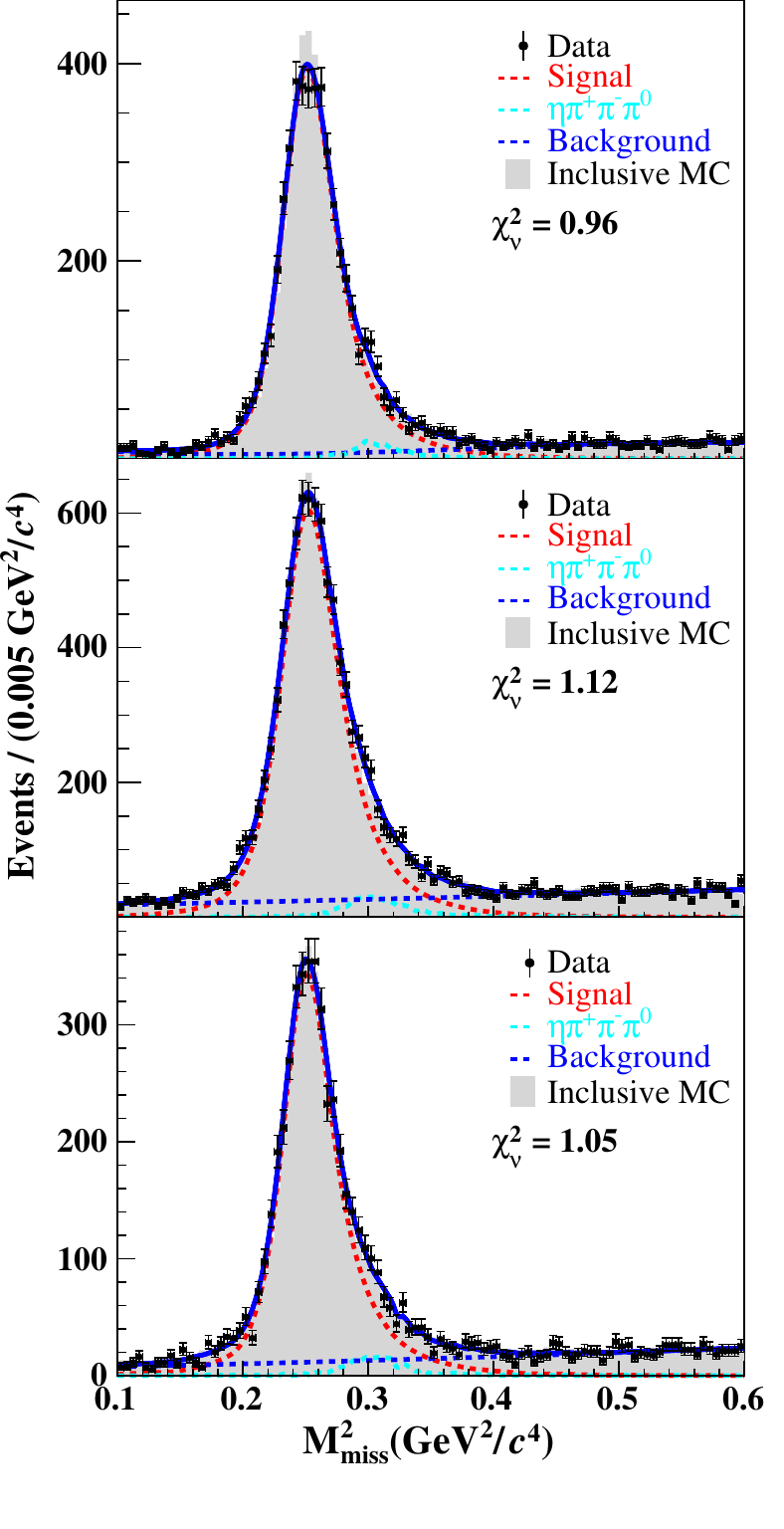}}
        \subfigure{\includegraphics[width=0.4\linewidth]{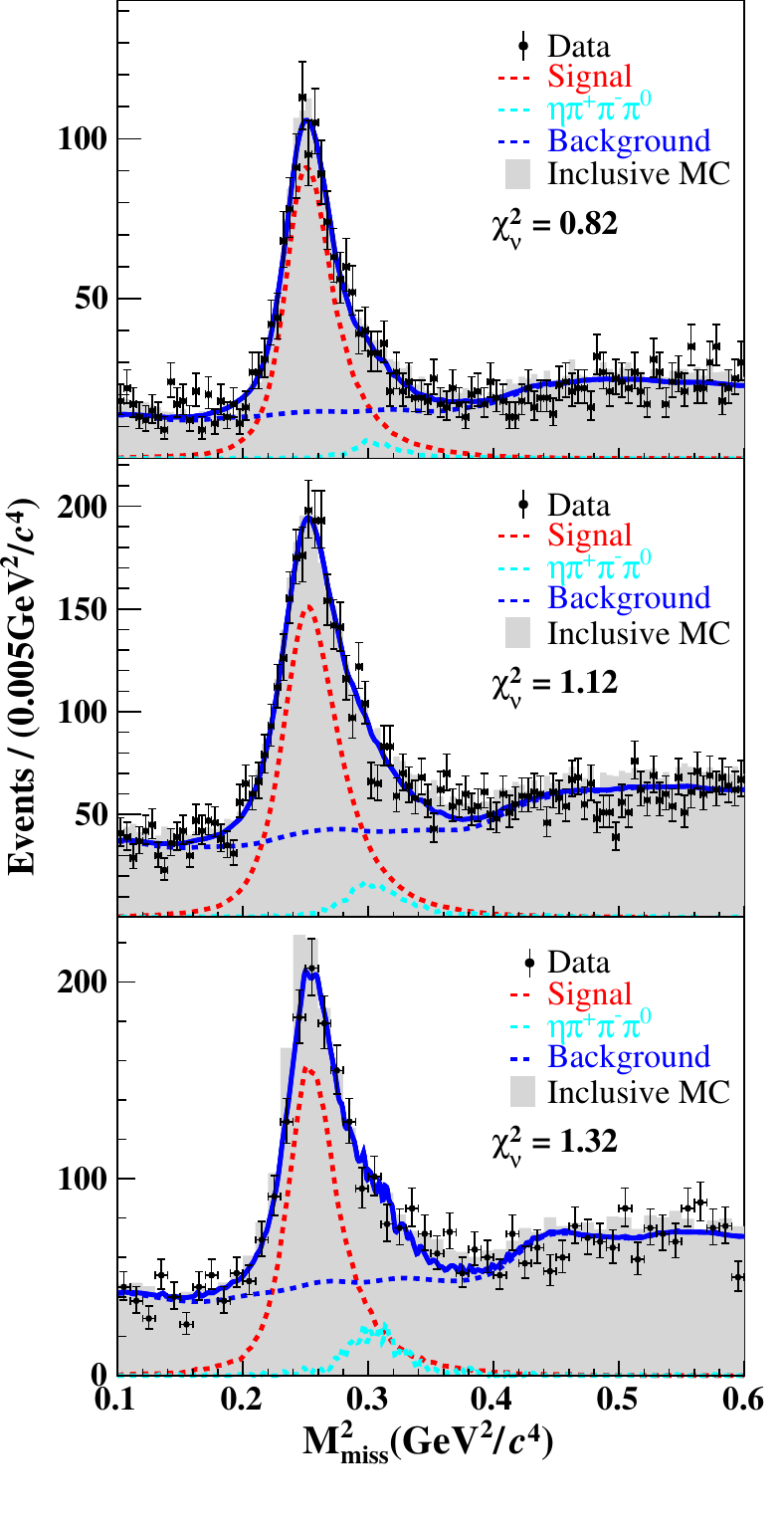}}        \abovecaptionskip=-10pt 
		\caption{Fit projections of the $\mmiss^2$ distributions for the candidates within the $\omega$ (left) signal and (right) sideband regions for the ST modes: (top) $\DzbartoKp$, (middle) $\DzbartoKpp$, and (bottom) $\DzbartoKppp$, individually. The dots with error bars are data, and the grey histograms are the inclusive MC simulation. The solid blue curves are the fit results, the dashed red, cyan and blue curves depict the signal shapes, the $\dz\to\eta\ppp$ background shapes, and the remaining background. The goodness of fit is quantified by the reduced chi-square, $\chi^2_\nu=\chi^2/{\rm n.d.f.}$, where the ${\rm n.d.f.} = 92, 95$ are numbers of degrees of freedom of the fits for the three ST modes within the $\omega$ (left) signal and (right) sideband regions.}
		\label{fig:mm2fitsig}
	\end{center}
\end{figure*}

In the above fits, the signal yields, $\NSG$ and $\NSB$, include contributions from $\dz\to\kshort\omega$, which must be subtracted.
The corresponding yields of $\dz\to\kshort\omega$ in the $\omega$ signal region ($\NSG_{\it\kshort}$) and sideband region ($\NSB_{\it\kshort}$) are estimated with the simulated MC sample, using the BF obtained in this analysis, as summarized in Table~\ref{tab:klres}.
The net signal yield $\NNET$ is calculated with 
\begin{equation}
{\NNET} = {\NSG}-{\NSG_{\it\kshort}} - f_{\rm scal} \, (\NSB-\NSB_{\it\kshort}),
\end{equation}
as summarized in Table~\ref{tab:klres}.

\begin{table}[htpb]
	\centering
	\caption{Summary of the yields $\NSG$, $\NSB$, $\NSG_{\it\kshort}$, $\NSB_{\it\kshort}$, $\NNET$, the scale factors $f_{\rm scal}$, and efficiencies $\epsilon^{\rm DT}$ for the three ST modes of the signal decay $D^0 \to \klong\omega$, where the uncertainties are statistical only.}
	\begin{tabular}{ c|c c c }
		\hline
		\hline
		ST mode & ~~~$\kaonp \pim$~~~ & ~~~$\kaonp \pim\piz$~~~ & ~~~$\kaonp \pim\pp$~~~ \\ \hline
		\hline
        $\NSG$ & $4916.2 \pm 93.8$ & $8188.3 \pm 130.7$& $4242.3 \pm 85.7$\\
        $\NSB$ & $1032.3 \pm 49.0$ & $1947.2 \pm 70.2$ & $929.2 \pm 45.7$\\ 
        $f_{\rm scal}$ & $0.378 \pm 0.027$ & $0.433 \pm 0.015$ & $0.415 \pm 0.030$\\
        \hline 
        $\NSG_{\kshort}$& $133.5 \pm 13.1$ & $259.6 \pm 18.4$& $142.4 \pm 13.5$\\
        $\NSB_{\kshort}$& $3.6 \pm 0.8$ & $14.4 \pm 1.5$& $4.3 \pm 0.9$\\ \hline

             $\NNET$& $4394.2\pm 99.5$& $7091.8 \pm 137.1$ & $3716.1 \pm 92.0$  \\
             $\epsilon^{\rm DT}(\%)$& $17.38 \pm 0.04$& $8.69 \pm 0.03$& $7.94 \pm 0.03$ \\
		\hline
		\hline
	\end{tabular}
	\label{tab:klres}
\end{table}

In addition to the above discussed backgrounds, there is potential background for the $\klong$ peaks from the non-$\DzDzbar$ events.
In principle, this background yield can be evaluated using events in the $\Mbc^{\rm tag}$ sideband region.
However, MC studies indicate that the $\mmiss^2$ distribution is shifted and has a worse resolution in the $\Mbc^{\rm tag}$ sideband region, since in Eq.~(\ref{eq3}), the energy and momentum of the $\klong\ppp$ system are replaced with the beam energy ($\Ebeam$) and the expected momentum of $\psi(3770)\to\DzDzbar$ ($\it \vec{P}_{\it\dz}$). 
To estimate the $\klong$ signal yield in the $\Mbc^{\rm tag}$ sideband region, an alternative calculation of $\mmiss^2$ is performed by replacing  $\Ebeam$ and $\it \vec{P}_{\it\dz}$ with the energy and momentum calculated from the kinematics of ST $\dzbar$ and the total energy and momentum.
MC studies validate that the $\klong$ signal is well-reconstructed in both the $\Mbc^{\rm tag}$ signal and sideband regions, though with slightly worse resolution.
The $\klong$ signal yields are extracted in the $\Mbc^{\rm tag}$ sideband and signal regions for the inclusive non-$\DzDzbar$ MC samples as well as in the $\Mbc^{\rm tag}$ sideband region for data. 
The $\klong$ signal yield in the $\Mbc^{\rm tag}$ signal region for the non-$\DzDzbar$ component is estimated to be $20.5\pm 8.4$, which is only 0.1\% of the total signal yield; therefore, this contribution is ignored in nominal results.

\section{Measurement of Branching Fractions}
\label{BFMEASURE}

Since the $\DzDzbar$ pair is produced coherently in $\psi(3770)$ decay, quantum correlations between the $\dz$ and $\dzbar$ needs to be considered. For each ST mode, $g$, and signal mode, $f$, the following relations hold~\cite{Ablikim_2024_xy}:

\begin{equation}
\begin{aligned}
	N^{\rm ST}_{g} & = 2N_{D^{0}\bar{D}^{0}} \, \mathcal{B}_{g} \, \epsilon^{\rm ST}_{g} \, (1+y_{D}^{2})\\
    & (1+r_{g}^{2}-2r_{g}R_{g} \, y_{D} \, \cos \delta_{g}),
  \end{aligned}
\end{equation}
\begin{equation}
\begin{aligned}
	N^{\rm DT}_{fg} & = 2N_{D^{0}\bar{D}^{0}} \, \mathcal{B}_{g} \, \mathcal{B}_{f} \, \mathcal{B}_{\rm sub} \, \epsilon^{\rm DT}_{fg} \, (1+y_{D}^{2})\\
    & [1+r_{g}^{2}-2r_{g}R_{g} \, \cos \delta_{g}(2F_{+}^{f}-1)]\;,
    \end{aligned}
\end{equation}

$\mathcal{B}_{g}$  ($\mathcal{B}_{f}$) is the BF of the tag (signal) mode, ${\it N}^{\rm DT}_{fg}$ and ${\it N}^{\rm ST}_{g}$ represent the numbers of DT and ST events, $N_{D^{0}\bar{D}^{0}}$ is the total number of $\DDbar$ pairs produced from $\ee$ collisions, and $\epsilon^{\rm DT}_{fg}$ and $\epsilon^{\rm ST}_{g}$ are the corresponding DT and ST efficiencies. To account for the reconstruction of particles through subsequent decays, the factor $\mathcal{B}^{\rm sub}_{f}$ is introduced, $\mathcal{B}^{\rm sub}_{f}=\mathcal{B}(\omega\to\ppp)\mathcal{B}(\piz\to\gamma\gamma){B}(\kshort\to\pp)$ for $\dz\to\kshort\omega$ and $\mathcal{B}^{\rm sub}_{f}=\mathcal{B}(\omega\to\ppp)\mathcal{B}(\piz\to\gamma\gamma)$ for for $\dz\to\klong\omega$.

Additionally, $y_{D}$ is the $\dz-\dzbar$ mixing parameter, $r_{g}^{2}$ is the ratio of the corresponding DCS amplitude to the CF amplitude, ${\it R}_g$ is the coherence factor ($0 \leq {\it R}_g \leq 1$) that quantifies the dilution from integrating over the phase space, $\delta_{g}$ is the strong-phase difference between these DCS and CF amplitudes, and ${\it F}_{+}^{f}= \rm{1 \, (0)}$ is the CP-even fraction of the $\dz\to \kshort\omega \, (\klong\omega)$ signal modes.
Ignoring the effects of $CP$ violation and assuming  $(1+r_{g}^{2}-2r_{g}R_{g} \, y_{D} \, \cos\delta_{g})\approx (1+r_{g}^{2})$, the absolute BFs of the self-conjugated signal mode $f$ incorporating quantum-correlation correction is given by
\begin{equation}
\begin{aligned}
	\mathcal{B}_{f}= \frac{ \Sigma_{g} {\it N}^{\rm DT}_{fg}}{\mathcal{B}^{\rm sub}_{f} \, \Sigma_{g}{\it N}^{\rm ST}_{g} \, \left( \frac{\epsilon^{\rm DT}_{fg}}{\epsilon^{\rm ST}_{g}} \right) \, \left[ 1 - \frac{2r_{g} {\it R}_{g} \cos\delta_{g}}{1+r_{g}^{2}}(2{\it F}_{+}^{f}-1) \right]}.
	\label{eq:brcal}
    \end{aligned}
\end{equation}
\noindent

Using Eq.~(\ref{eq:brcal}), the BFs for $\dz \to \kshort\omega$ and $\dz\to\klong \omega$ are calculated for the three ST modes individually and for their average, as summarized in Table~\ref{tab:KLBR}. 
These results incorporate the ${\it N}^{\rm DT}_{fg}$ and ${\it N}^{\rm ST}_{g}$ and $\epsilon^{\rm DT}_{fg}$ and $\epsilon^{\rm ST}_{g}$ in 

Tables~\ref{tab:ksres} and \ref{tab:klres}, the decay parameters of the $\dz$ ($r^{2}$,${\it R}$, $\delta$) in PDG~\cite{pdg}, as well as the BFs of $\omega\to\ppp$, $\piz\to\gamma\gamma$ and $\kshort\to\pp$ in the PDG~\cite{pdg},

Reasonable agreement between the resultant BFs is seen. Therefore, the average values of BFs by combining the three ST modes are obtained.
According to Eq.~(\ref{eq:Rdef}), the $\kshort$-$\klong$ asymmetry is determined to be  $R(\dz, \ksl\omega)=(4.2\pm 1.0)\%$, where the uncertainty is statistical only.

\begin{table}[H]
	\centering
	\caption{Absolute BFs ($\times 10^{-3}$) of $\dz \to \kshort \omega$ and $\dz \to \klong \omega$, calculated using the three ST modes individually and their average; uncertainties are statistical only.}
	\begin{tabular}{ l|c c }
		\hline
		\hline
		Tag mode & ~~~$\mathcal{B}(\dz \to \kshort \omega)$~~~ & ~~~$\mathcal{B}(\dz \to\klong \omega)$~~~ \\
		\hline
        $\DzbartoKp  $ & $12.04 \pm 0.35$ & $11.51 \pm 0.26$ \\
        $\DzbartoKpp $ & $11.64 \pm 0.27$ & $10.68 \pm 0.20$ \\
        $\DzbartoKppp$ & $11.84 \pm 0.36$ & $10.86 \pm 0.28$ \\ \hline
        Average        & $11.79 \pm 0.19$ & $10.84 \pm 0.14$ \\

		\hline
		\hline
	\end{tabular}
	\label{tab:KLBR}
\end{table}

\section{Systematic uncertainties}
\label{sec:sys}
In this analysis, the DT method cancels most of the systematic uncertainties associated with the ST selection. 
The relative systematic uncertainties in the BF measurement are addressed below in detail.

The systematic uncertainties associated with the tracking of the charged pions are studied using a control sample of DT $\psi(3770)\to D^0\bar{D}^0(D^+\bar{D}^-)$. 

The MC samples are weighted by the efficiency ratio between data and MC simulation as a function of charged particle momentum.
The differences between re-weighted and nominal results, $1.0\%$ for $\dz \to \kshort \omega $ and $0.8\%$ for $\dz \to \klong \omega$ are assigned as systematic uncertainties. 
The systematic uncertainties due to the PID are estimated using a similar method, resulting in an uncertainty of $0.2\%$ for both decay modes.
The systematic uncertainties arising from the $\piz$ reconstruction are studied as a function of $\piz$ momentum using a control sample of $D^0\to K^-\pip\piz$, and are determined through a similar re-weighting of the data-MC differences. The uncertainties related to the $\piz$ reconstruction are assigned to be $1.1\%$ for both decay modes. 

In the analysis of $\dz\to\kshort \omega$, the systematic uncertainty associated with the $\kshort$ reconstruction is studied based on the control samples of $\dz\to\kshort\pp, \dz\to\kshort\ppp,$ and $\dz\to\kshort\piz$, a similar re-weighting of the data-MC differences is determined, and the difference between re-weighted and nominal results, $1.1\%$, is assigned as the systematic uncertainty.
The systematic uncertainty associated with the $\DeltaE^{\rm sig}$ requirement is studied by smearing the signal MC sample with a Gaussian function to better match the distribution in data; the resultant difference of the efficiency, 0.05\%, is negligible. Here, the parameters of the Gaussian function are obtained by fitting the $\DeltaE^{\rm sig}$ distribution of data with a MC simulated shape convolved with the Gaussian function. 

In the analysis of $\dz\to\klong \omega$, the systematic uncertainties associated with the requirements on extra charged tracks and the total energy of extra showers are studied with a DT control sample of $\dzbar \to \klong\ppp$ with an additional requirement of $\Mppp<$ 0.64$\gevcc$ or $\Mppp >$ 0.92$\gevcc$ to veto the $\dz\to\klong\omega$ signal. The differences of data and MC efficiencies are $(-0.75\pm0.15)\%$ for the no extra charged tracks and $(-0.22\pm0.18)\%$ for the $\it E_{\gamma}^{\text{extra}}$ requirement. The  systematic uncertainties are assigned to be $0.9\%$ and $0.4\%$, respectively. 

The systematic uncertainty associated with the ST yield is estimated to be $0.3\%$, derived from the alternative fits of the $\Mbc^{\rm tag}$ distributions by varying the endpoint of Argus function of the background shape from 1.8865~$\gevcc$ to 1.8864~$\gevcc$ or 1.8866~$\gevcc$ , and varying the parameters of the mean values and sigma values of two Gaussian functions  of the signal shape by $\pm 1\sigma$.

The systematic uncertainties related to the DT yields are estimated by varying the fitting range, signal and background shapes used in the DT fits, and the estimation of peaking backgrounds. For $\dz \to \kshort\omega$, the systematic uncertainty is determined by varying the fitting range by $\pm0.005\gevcc$, varying the means and widths of two Gaussian functions by $\pm 1\sigma$ of signal, varying the fixed parameters in ARGUS function by $\pm 1\sigma$ of BKGI and BKGII, changing the Student's $t$-function to a bifurcated Student's $t$-function with different $n$ and $\sigma$ on the left and right sides of the maximum value and two ARGUS functions to one ARGUS function for the ($\Mbc^{\rm tag} +\Mbc^{\rm sig}$) dimension of BKGIII, and convolving the background shapes of BKGIV and BKGV with Gaussian function. The quadrature sums of the relative changes of the DT yields for the individual changes is 0.5\%, which is assigned as a systematic uncertainty.
For $\dz \to\klong \omega$, a systematic uncertainty of $0.7\%$ is adopted, determined by varying the fitting range by $\pm0.015\gevcc$, varying the truth-match angle requirement by $\pm5^\circ$ of signal and $\dz\to\eta\ppp$ background, varying the order of Chebyshev polynomial function of combinatorial backgrounds, and varying the yields of peaking background $\dz\to\kshort\omega$ by $\pm1\sigma$.

The uncertainties associated with the $\omega$ mass window arise from the resolution difference between data and MC simulation. Fits are performed on the $\Mppp$ distributions, and the difference in the ratios of $\omega$ yields inside and outside the signal region between data and MC is taken as the uncertainty, which is 0.1\% for both $\dz \to \kshort\omega $ and $\dz \to \klong\omega$.
The uncertainties associated with the quantum correlation~(QC) input parameters $r$, $R$ and $\delta$ are estimated by varying their values by their uncertainties in the BF calculation, and the largest resultant deviations from the nominal value, $0.6\%$ for $\dz \to \kshort \omega$ and $0.5\%$ for $\dz \to \klong \omega$, are taken as uncertainties. 
The uncertainties from MC simulation sample size used to determine detection efficiencies are $0.3\%$. 
The uncertainty associated with intermediate-decay BFs is 0.7\%.

All the above uncertainties are summarized in Table~\ref{tab:totsyst}.
Assuming the sources are independent, the total uncertainties are their sums in quadrature. 
\begin{table}[htb]
	\centering
	\caption{Systematic uncertainties, in \%, on the BF measurements.}
	\begin{tabular}{ c|c|c }
		\hline
		\hline
		Source & ~~~$\dz \to \kshort \omega$~~~ & ~~~$\dz \to \klong \omega$~~~\\
		\hline
		$\pi^{\pm}$ tracking            & 1.0 & 0.8 \\
		$\pi^{\pm}$ PID                 & 0.2 & 0.2 \\
		$\piz$ reconstruction           & 1.1 & 1.1 \\
        $\kshort$ reconstruction        & 1.1 & - \\
        $N_{\text{charged}}^{\text{extra}} $ requirement & - & 0.9 \\
		$E_{\gamma}^{\text{extra}}$ requirement & - & 0.4 \\
        ST yields             & 0.3 & 0.3 \\	
		DT yields             & 0.5 & 0.7 \\
        $\omega$ mass window  & 0.1 & 0.1 \\ 
		QC correction        & 0.6 & 0.5 \\
		MC statistics & 0.3 & 0.3 \\
		Daughter BFs & 0.7 & 0.7 \\
		\hline
		Total & 2.2 & 2.1 \\
		\hline
		\hline
	\end{tabular}
	\label{tab:totsyst}
\end{table}

Regarding the systematic uncertainties associated with the measurement of ${ R(\dz,\ksl \omega)}$, the selection criteria for the $\omega$ and ST sides are identical for both $\dz \to \kshort\omega$ and $\dz \to \klong\omega$. Consequently, the uncertainties related to $\pi^{\pm}$ tracking and PID, $\piz$ reconstruction,  QC corrections, $\omega$ mass window, and the quoted BFs \(\mathcal{B}(\omega \to \ppp)\) and  \(\mathcal{B}(\piz \to \gamma\gamma)\) cancel.
The remaining contributions lead to a systematic uncertainty of 1.2\% on the measurement of ${ R(\dz,\ksl \omega)}$.

In this analysis, signal yields are extracted by fitting specific distributions, and the interference between processes with and without the $\omega$ intermediate state are not considered.
Based on the inclusive MC, the fraction of $\dz\to\omega\kshort$ in $\dz\to\kshort\ppp$ is $19.1\%$. After removing the $\dz\to\omega\kshort$ components, without influence from interference between $\omega$ and non-$\omega$ components, the fraction of new and original total amplitudes is $81.6\%$, and the fraction of $\dz\to\omega\kshort$ in $\dz\to\kshort\ppp$, without $\omega$ and non-$\omega$ component interference, is estimated to be $18.4\%$
The MC study indicates that this interference will affect the BFs of $\dz\to\kshort/\klong \omega$ by up to 4.0\%, which is taken as the third uncertainties in the BF measurement.

\section{SUMMARY AND DISCUSSION}
\label{sum}

\begin{table*}[htbp]
	\centering
	\caption{Comparisons of experimental measurements and theoretical calculations for the  BFs of $D^0\to K^0_{S,L}\omega$ and and the $K_{S}^{0} - K_{L}^{0}$ asymmetry. The first uncertainties are statistical, the second systematic, and the third from interference between $\dz \to \ksl\omega$ and non-resonant $\dz \to \ppp\ksl$ processes.}
	\begin{tabular}{ c|c|c|c }
		\hline
		\hline
		Result & $\mathcal{B}(D^{0} \to K_{S}^{0} \omega)$ & $\mathcal{B}(D^{0} \to K_{L}^{0} \omega)$ & $R(D^{0}, \ksl\omega)$ \\
		\hline
		This work & $(11.79 \pm 0.19 \pm 0.26 \pm 0.47)\times 10^{-3}$ & $(10.84 \pm 0.14 \pm 0.23 \pm 0.44) \times 10^{-3}$ & $(4.2 \pm 1.0 \pm 0.9 \pm 2.8)\%$ \\ 
        Previous work & $(11.20\pm 0.40\pm0.50)\times 10^{-3}$~\cite{Brksomega}  & $(11.64 \pm 0.22 \pm 0.28)\times 10^{-3}$~\cite{KLX}  & $(-2.4 \pm 3.1)\%$ ~\cite{KLX} \\
        \hline
		FAT~\cite{PRD.95.073007} & $(11.8 \pm 1.9) \times 10^{-3}$ & $(9.5 \pm 1.5) \times 10^{-3}$ & $(11.3 \pm 0.1)\%$ \\
		F4~\cite{PRD.109.073008} & $(12.7 \pm 0.6) \times 10^{-3}$ & $(10.3 \pm 0.5) \times 10^{-3}$ & $(10.6 \pm 3.4)\%$ \\
		F1$^\prime$~\cite{PRD.109.073008} & $(12.5 \pm 0.6) \times 10^{-3}$ & $(10.5 \pm 0.5) \times 10^{-3}$ & $(8.9 \pm 3.5)\%$ \\
		\hline
		\hline
	\end{tabular}
	\label{tab:sum}
\end{table*}
Based on 7.93~$\ifb$  of $\ee$ annihilation data with a center-of-mass energy of 3.773~$\gev$ collected by the BESIII detector, we perform the measurement of the absolute BFs of $\dz\to\kshort \omega$ and $\dz\to\klong \omega$ using a DT method, and extract the $\kshort$-$\klong$ asymmetry. The results are 

summarized in Table~\ref{tab:sum}, and compared with the previous measurements as well as theoretical predictions from different models.

By considering the statistical and systematic uncertainties, the measured BF of $\dz\to\kshort\omega$ is consistent with that reported by the CLEO collaboration~\cite{Brksomega} with a precision improved by a factor of 2.
The BF of $\dz\to\klong\omega$ is consistent with our previous measurement~\cite{KLX} using an integrated luminosity of 2.93~\ifb, with precision improved by a factor of 1.3.
Notably, compared to the previous experiments, the BF of $\dz\to\kshort\omega$ increases, while that of $\dz\to\klong\omega$ decreases, and they are closer to the mean values of different theoretical predictions~\cite{PRD.95.073007,PRD.109.073008}.
Consequently, the value of $R(D^{0}, \ksl\omega)$ increases compared to our previous measurement ~\cite{KLX}.
The value of $R(D^{0}, \ksl\omega)$ is of the same sign and closer to those of theoretical predictions.  
These results are valuable for accessing individual DCS amplitudes involving $K^0$, which can only be measured with quantum correlated $\ee\to\DDbar$  production near the threshold. Therefore, they provide further understanding the decay mechanism and flavor SU(3) symmetry and the $\dz - \dzbar$ mixing~\cite{PRD.81.114020, PRD.80.076008, PRD.86.014014, PRD.92.014004, PRD.95.073007, CPC.42.063101, PRD.109.073008}.

Our larger statistics leads to smaller statistical and systematic uncertainties of the measurements as expected. However, the effects of interference between $\dz\to\ksl \omega$ and $\dz\to\ksl \ppp$ on the BFs and $\klong$-$\kshort$ asymmetry $R(D^{0}, \ksl\omega)$ must be considered, as has been done here for the first time. This increases the total uncertainty compared to earlier measurements. Therefore, with larger data samples currently, amplitude analysis techniques must be applied to further improve the precision.

\section{ACKNOWLEDGEMENT}
 \input{acknowledgement_2025-11-25}

\bibliography{main}
	
\end{document}

%% file: authorlist_2025-11-25.tex
M.~Ablikim$^{1}$\BESIIIorcid{0000-0002-3935-619X},
M.~N.~Achasov$^{4,d}$\BESIIIorcid{0000-0002-9400-8622},
P.~Adlarson$^{82}$\BESIIIorcid{0000-0001-6280-3851},
X.~C.~Ai$^{88}$\BESIIIorcid{0000-0003-3856-2415},
C.~S.~Akondi$^{31A,31B}$\BESIIIorcid{0000-0001-6303-5217},
R.~Aliberti$^{39}$\BESIIIorcid{0000-0003-3500-4012},
A.~Amoroso$^{81A,81C}$\BESIIIorcid{0000-0002-3095-8610},
Q.~An$^{78,64,\dagger}$,
Y.~H.~An$^{88}$\BESIIIorcid{0009-0008-3419-0849},
Y.~Bai$^{62}$\BESIIIorcid{0000-0001-6593-5665}
% ...
\\
\vspace{0.2cm}
(BESIII Collaboration)\\
\vspace{0.2cm} {\it
$^{1}$ Institute of High Energy Physics, Beijing 100049, People's Republic of China\\
% ...
$^{4}$ Budker Institute of Nuclear Physics SB RAS (BINP), Novosibirsk 630090, Russia\\
% ...
$^{31}$ INFN Sezione di Ferrara, (A)INFN Sezione di Ferrara, I-44122, Ferrara, Italy; (B)University of Ferrara, I-44122, Ferrara, Italy\\
% ...
$^{39}$ Johannes Gutenberg University of Mainz, Johann-Joachim-Becher-Weg 45, D-55099 Mainz, Germany\\
% ...
$^{62}$ Southeast University, Nanjing 211100, People's Republic of China\\
% ...
$^{64}$ State Key Laboratory of Particle Detection and Electronics, Beijing 100049, Hefei 230026, People's Republic of China\\
% ...
$^{78}$ University of Science and Technology of China, Hefei 230026, People's Republic of China\\
% ...
$^{81}$ University of Turin and INFN, (A)University of Turin, I-10125, Turin, Italy; (B)University of Eastern Piedmont, I-15121, Alessandria, Italy; (C)INFN, I-10125, Turin, Italy\\
$^{82}$ Uppsala University, Box 516, SE-75120 Uppsala, Sweden\\
% ...
$^{88}$ Zhengzhou University, Zhengzhou 450001, People's Republic of China\\

\vspace{0.2cm}
$^{\dagger}$ Deceased\\
% ...
$^{d}$ Also at the Novosibirsk State University, Novosibirsk, 630090, Russia\\
% ...
}

%% file: acknowledgement_2025-11-25.tex

The BESIII Collaboration thanks the staff of BEPCII (https://cstr.cn/31109.02.BEPC) and the IHEP computing center and the supercomputing center of the University of Science and Technology of China (USTC) for their strong support. This work is supported in part by National Key R\&D Program of China under Contracts Nos. 2023YFA1606000, 2023YFA1606704, 2023YFA1609400; National Natural Science Foundation of China (NSFC) under Contracts Nos. 11635010, 11935015, 11935016, 11935018, 12025502, 12035009, 12035013, 12061131003, 12192260, 12192261, 12192262, 12192263, 12192264, 12192265, 12221005, 12225509, 12235017, 12342502, 12361141819, 12122509, 12105276; the Chinese Academy of Sciences (CAS) Large-Scale Scientific Facility Program; CAS Youth Team Program under Contract No. YSBR-101; Joint Large-Scale Scientific Facility Funds of the NSFC and CAS under Contracts Nos. U2032111; the Strategic Priority Research Program of Chinese Academy of Sciences under Contract No. XDA0480600; CAS under Contract No. YSBR-101; 100 Talents Program of CAS; The Institute of Nuclear and Particle Physics (INPAC) and Shanghai Key Laboratory for Particle Physics and Cosmology; ERC under Contract No. 758462; German Research Foundation DFG under Contract No. FOR5327; Istituto Nazionale di Fisica Nucleare, Italy; Knut and Alice Wallenberg Foundation under Contracts Nos. 2021.0174, 2021.0299, 2023.0315; Ministry of Development of Turkey under Contract No. DPT2006K-120470; National Research Foundation of Korea under Contract No. NRF-2022R1A2C1092335; National Science and Technology fund of Mongolia; Polish National Science Centre under Contract No. 2024/53/B/ST2/00975; STFC (United Kingdom); Swedish Research Council under Contract No. 2019.04595; U. S. Department of Energy under Contract No. DE-FG02-05ER41374
